\documentclass{article}

  \usepackage[nonatbib,preprint]{neurips_2024}

\usepackage{booktabs}
\usepackage{makecell}
\usepackage{multirow}
\usepackage{graphicx} % Required for inserting images
\usepackage{xspace}
\usepackage[skins]{tcolorbox}
\usepackage{caption}
\usepackage{threeparttable}
\usepackage{changepage}
\usepackage{wrapfig}
\usepackage{amsmath}
\usepackage{listings}
\usepackage[colorlinks=true, linkcolor=blue,  citecolor=black, urlcolor=blue]{hyperref}

\usepackage{enumitem}

\newcommand{\tool}[0]{\textsc{DepsRAG}\xspace}

\newcounter{promptbox}

\newcommand{\dependencygraphagent}{\texttt{DependencyGraphAgent}\xspace}
\newcommand{\searchagent}{\texttt{SearchAgent}\xspace}
\newcommand{\assistantagent}{\texttt{AssistantAgent}\xspace}
\newcommand{\criticagent}{\texttt{CriticAgent}\xspace}

\newcommand{\prompt}[3][]{%
    \refstepcounter{promptbox} % Step the custom counter
    \begin{tcolorbox}[
        space to upper,
        colframe=black!80!white,
        colback=white,
        halign=left,
        valign=top,
        halign lower=flush left,
        title=#2,
        fonttitle=\bfseries,
    ]
    #3
    \end{tcolorbox}%
    \if\relax\detokenize{#1}\relax
    \else
        \phantomsection % Add a phantom section to create a hyper-reference target
        \label{#1}  % Label for referencing
    \fi
}

\usepackage{listings}
\usepackage{xcolor}

\title{\tool: Towards Agentic Reasoning and Planning for Software Dependency Management}

\author{%
  Mohannad Alhanahnah \\
  University of Wisconsin-Madison, USA \\
  \texttt{mohannad@cs.wisc.edu} \\
  \And
  Yazan Boshmaf \\
  Qatar Computing Research Institute, Qatar \\
  \texttt{yboshmaf@hbku.edu.qa} \\
}

\begin{document}
\maketitle

\begin{abstract}
In the era of Large Language Models (LLMs) with their advanced capabilities, a unique opportunity arises to develop LLM-based digital assistant tools that can support software developers by facilitating comprehensive reasoning about software dependencies and open-source libraries before importing them. This reasoning process is daunting, mandating multiple specialized tools and dedicated expertise, each focusing on distinct aspects (e.g., security analysis tools may overlook design flaws such as circular dependencies, which hinder software maintainability). Creating a significant bottleneck in the software development lifecycle. 

In this paper, we introduce \tool, a multi-agent framework designed to assist developers in reasoning about software dependencies. \tool first constructs a comprehensive Knowledge Graph (KG) that includes both direct and transitive dependencies. Developers can interact with \tool through a conversational interface, posing queries about the dependencies. \tool employs Retrieval-Augmented Generation (RAG) to enhance these queries by retrieving relevant information from the KG as well as external sources, such as the Web and vulnerability databases, thus demonstrating its adaptability to novel scenarios. \tool incorporates a Critic-Agent feedback loop to ensure the accuracy and clarity of LLM-generated responses. We evaluated \tool using GPT-4-Turbo and Llama-3 on three multi-step reasoning tasks, observing a threefold increase in accuracy with the integration of the Critic-Agent mechanism. \tool demo and implementation are available: \url{https://github.com/Mohannadcse/DepsRAG}.

\end{abstract}

% \vspace{-0.2cm}
\section{Introduction}

The increasing reliance on code reuse in modern software development, facilitated by third-party packages, has substantially enhanced both software quality and productivity~\cite{productivity}. As the adoption of open-source and third-party libraries becomes more widespread, securing the software supply chain has emerged as a critical concern~\cite{rack2023jack}. In response, regulatory measures such as Executive Order 14028~\cite{nistExecutiveOrder} and the 2023 EU Cyber Resilience Act (CRA)~\cite{CRA} impose strict requirements on developers to ensure compliance of software dependencies, including open-source and third-party libraries, within their projects.
A recent survey~\cite{tidelift2022Open} highlights a widespread lack of confidence among developers in the effectiveness of their current dependency management practices. This uncertainty contributes to a prolonged and cumbersome approval process for incorporating new open-source libraries for 61\% organizations, thus introducing a critical bottleneck in the software development lifecycle.
This challenge is further compounded by the lack of advanced tools capable of thoroughly analyzing and addressing security and maintainability concerns in open-source and third-party libraries. Existing security tools often overlook essential issues, such as detecting circular dependencies or resolving version conflicts. For example, within the Python ecosystem, developers typically rely on tools such as ``pip-audit''~\footnote{https://pypi.org/project/pip-audit/} to perform security checks and ``piptree''~\footnote{https://pypi.org/project/piptree/} to visualize dependencies in a hierarchical format. However, detecting circular dependencies often requires manual inspection or the use of additional tools such as ``pipdeptree''~\footnote{https://pypi.org/project/pipdeptree/}.

To tackle these challenges, we introduce \tool, an LLM-powered digital assistant designed to facilitate informed decision making on software dependencies before their integration. Central to \tool’s architecture is a multi-agent system that incorporates Retrieval-Augmented Generation (RAG) alongside planning and critique agents, enabling in-depth reasoning about direct and transitive dependencies. These dependencies are modeled as a Knowledge Graph (KG). The developer can then interact with \tool to answer questions such as the number of dependencies, the identification of the key packages, the depth of the graph, and the dependency paths. Furthermore, \tool is capable of acquiring knowledge from external sources, including the Web and vulnerability databases, to respond to unforeseen queries.

In contrast to conventional tools that depend on predefined queries, \tool uses large language models (LLMs) to dynamically generate queries, significantly enhancing scalability and adaptability. Using the RAG, \tool integrates KG data to improve the accuracy and relevance of responses to user questions. To further ensure the precision and reliability of its results, \tool uses a Critic-Agent interaction, which iteratively refines its responses through reasoning and validation, producing consistently accurate and reliable results.

\section{Background}
% This section provides an overview of some concepts and patterns that we relied on.
% \vspace{-0.15cm}
\subsection{Dependency Graphs}

Dependency graphs are essential in software engineering for representing dependencies among software entities like classes, functions, modules, packages, or larger components. Each node represents an entity and directed edges indicate dependencies, meaning changes in one entity can affect another.

Litzenberger et al. \cite{DGMF} proposed a framework with three levels of dependency granularity: package-to-package, artifact-to-package, and artifact-to-artifact. D\"{u}sing and Hermann \cite{vlunAnalysis} used artifact-to-package graphs to study vulnerability propagation in software repositories. Benelallam et al. \cite{Maven_depgraph} created the Maven Dependency Graph to explore artifact releases, evolution, and usage trends. Dependency graphs are also utilized in tools for program understanding \cite{falke2005dominance, musco2017generative}.

% \vspace{-0.2cm}
\subsection{Knowledge Graphs (KGs)}

Knowledge graphs (KGs) organize information in a structured format, capturing relationships between real-world entities and making them comprehensible to humans and machines~\cite{agrawal2023can}. In a KG, data are organized as triplets (head entity, relation, tail entity), where the relation is the relationship between these two entities, such as ("Steven Jobs", "owns", "Apple"). More formally, KGs store structured knowledge as a collection of triples $\mathcal{KG}=\{(e_i,r,e_j)\subseteq \mathcal{E}\times \mathcal{R}\times \mathcal{E}\}$, where $\mathcal{E}$ and $\mathcal{R}$ denote the set of entities and relations, respectively. KGs are created by describing entities and entity relationships, known as graph schema. KGs are useful for a variety of applications, such as question-answering~\cite{hao2017end}, information retrieval~\cite{xiong2017explicit}, recommender systems~\cite{zhang2016collaborative}, cybersecurity~\cite{agrawal2023aiseckg,codePropertyGraph}, and natural language processing~\cite{yang2017leveraging}.

\subsection{Agentic LLM Applications}
\textbf{Agents.} An agent is responsible for a specific aspect of a task, acting essentially as a message transformer. In code, an agent is typically represented as a class encapsulating an interface to an LLM, along with optional tools and external data (e.g., a vector or graph database). Agents communicate by exchanging messages, similar to the actor model in programming~\cite{actor}. For practical applications, agents can trigger actions (e.g., API calls, computations) and access external data, facilitated by tools and RAG.

\noindent
\textbf{Tools, also known as functions or plugins.} Tools allow LLMs to trigger external actions beyond generating text. While free-form text is useful for descriptions, summaries, or agent queries, structured outputs are needed for actions like API calls, code execution, or database queries. In such cases, the LLM produces a structured response, typically in JSON format, specifying necessary details like code or query parameters. The LLM uses a tool when generating such structured responses, and a tool handler is defined to execute the corresponding action when recognized in the LLM’s output.

\noindent
\textbf{Retrieval-Augmented Generation (RAG).}
Utilizing an LLM in isolation presents two primary limitations: (a) its responses are confined to the knowledge available at the time of pre-training, making it incapable of addressing questions related to private sources (such as databases or documents) or post-training information, and (b) it cannot verify the correctness of its responses. RAG mitigates these challenges by generating answers based on specific documents or data while also providing source references for validation. When an LLM receives a query $Q$ in the RAG process, it retrieves a set of $k$ most relevant pieces of data $D = {d_1, d_2, \dots, d_k}$ from a database store. The query is then reformulated into a new prompt: ''Given the following data: $[d_1, d_2, \dots, d_k]$, provide an answer to this question: $Q$, based ONLY on these data, and indicate which data support your answer.'' 

\noindent
\textbf{Multi-Agent Orchestration}
An orchestration mechanism is critical for ensuring task progression, managing message flow, and handling deviations from instructions. This work uses a multi-agent programming approach where agents communicate via message exchange for both inter-agent and intra-agent interactions. Frameworks like Langroid~\cite{langroid} and Langchain~\cite{langchain} provide robust tools for message routing and task delegation. In this work, Langroid is employed, where orchestration is encapsulated in the Task class. This class manages user interactions, tool integration, and sub-task delegation by processing a "current pending message" (CPM) through responders, iterating through steps until task completion.
\section{Motivating Example and Challenges}
\label{sec:motivation}
In this Section, we present the following critical software dependency task to identify risky dependencies. 

\begin{quote}
``\textit{For package X, version Y under software ecosystem Z, which packages in X have the most dependencies relying on them, and what is the risk associated with a vulnerability in those packages?}''
\end{quote}
% (i.e., which nodes have the highest in-degree in the graph)

We emphasize that developing an AI digital assistant system for managing software dependencies and answering tasks like the above example involves several key challenges: (a) the complex, hierarchical nature of software dependencies, which complicates the direct application of standard RAG methods; (b) the fragility of large language models (LLMs), prone to instruction deviation, hallucination, and inaccurate outputs; (c) decomposing the task into manageable sub-tasks while extracting relevant information from the dependency graph and assessing associated risks by retrieving vulnerability data; (d) aggregating data from various sources to formulate a comprehensive response; (e) efficiently routing and delegating tasks to appropriate agents, while mitigating risks like infinite loops and deadlocks through careful orchestration~\cite{MALADE}; and (f) validating the final response and refining the query to enhance retrieval accuracy from the knowledge graph (KG).

\begin{figure*}[ht]
    \centering    \includegraphics[width=\textwidth]{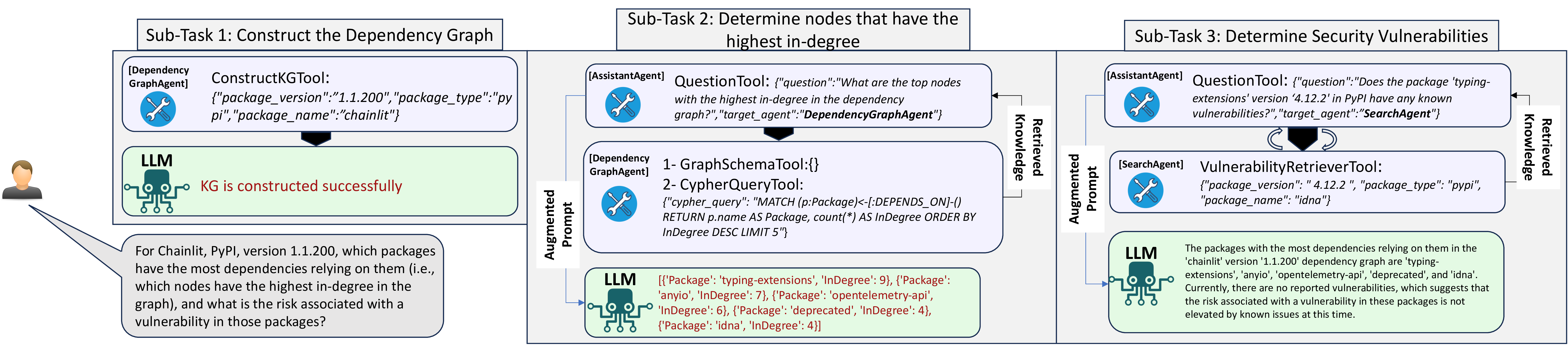}
    \caption{A real-world demonstration of \tool, applied to identify critical packages and associated risks within the dependency graph of the \texttt{Chainlit} package version 1.1.200. This process involves a sequence of subtasks executed by two agents: \dependencygraphagent and \searchagent. The question in Sub-Task 3 will be repeated for each package in the LLM response in Sub-Task 2.
    Therefore, \assistantagent needs to orchestrate the communication between these agents, by setting the field ``target\_agent'' in JSON message generated by \texttt{QuestionTool}, aggregate the responses received in Sub-Task 2 and Sub-Task 3, and subsequently forward the agents' responses to the \criticagent for validation (omitted in this Figure for the sake of simplicity).}
    \label{fig:steps}
\end{figure*}

Therefore, we designed \tool to handle these complex tasks by decomposing them into sub-tasks as depicted in Figure~\ref{fig:steps}. Accordingly, given a query of this form, \tool executes the following sub-tasks, given X, Y, and Z:

\begin{adjustwidth}{1.1cm}{}
\begin{enumerate}[label=Sub-Task {{\arabic*}}:]
    \item Construct the dependency graph for the software package X, version Y, and from ecosystem Z. The agent \dependencygraphagent will construct the dependency graph by executing the tool \texttt{ConstructKGTool}.
    \item Determine nodes that have the highest in-degree in the dependency graph. The agent \dependencygraphagent will translate this question into a query to retrieve from the dependency graph the highest in-degree nodes by executing the tools \texttt{GraphSchemaTool} and \texttt{CypherQueryTool}.
    \item Query the vulnerability database for each node obtained in STEP 2 and generate a structured report identifying vulnerable nodes within the dependency graph. The \searchagent leverages the \texttt{VulnerabilityTool} to query the vulnerability database, returning vulnerability results for each node to the \assistantagent, which subsequently consolidates the findings into a structured report. 
\end{enumerate}
\end{adjustwidth}

Both \dependencygraphagent and \searchagent can be paired with a \criticagent that evaluates and provides feedback on the outputs generated by the primary agent (i.e., \assistantagent). Based on this feedback, the primary agent iteratively refines its response. This Agent-Critic loop continues until the Critic agent endorses the final output. This interaction paradigm substantially improves the reliability of \tool. 
Section~\ref{sec:depsrag-arch} introduces the details of \tool and discusses in detail the Agent-Critic interaction.
\section{\tool: Multi-Agent System for Dependency Management}
\label{sec:depsrag-arch}
This section presents the architecture of \tool, a chatbot designed for the automated construction of software dependencies as a KG and the subsequent answering of user queries, expressed in natural language, regarding the dependency graph. 

\subsection{\tool Agent-Critic Interaction and Orchestration Architecture}
\label{sec:arch}

\begin{wrapfigure}{r}{0.5\linewidth}  % 'r' indicates right side wrapping, adjust width as needed
    \centering
    \includegraphics[width=\linewidth, keepaspectratio]{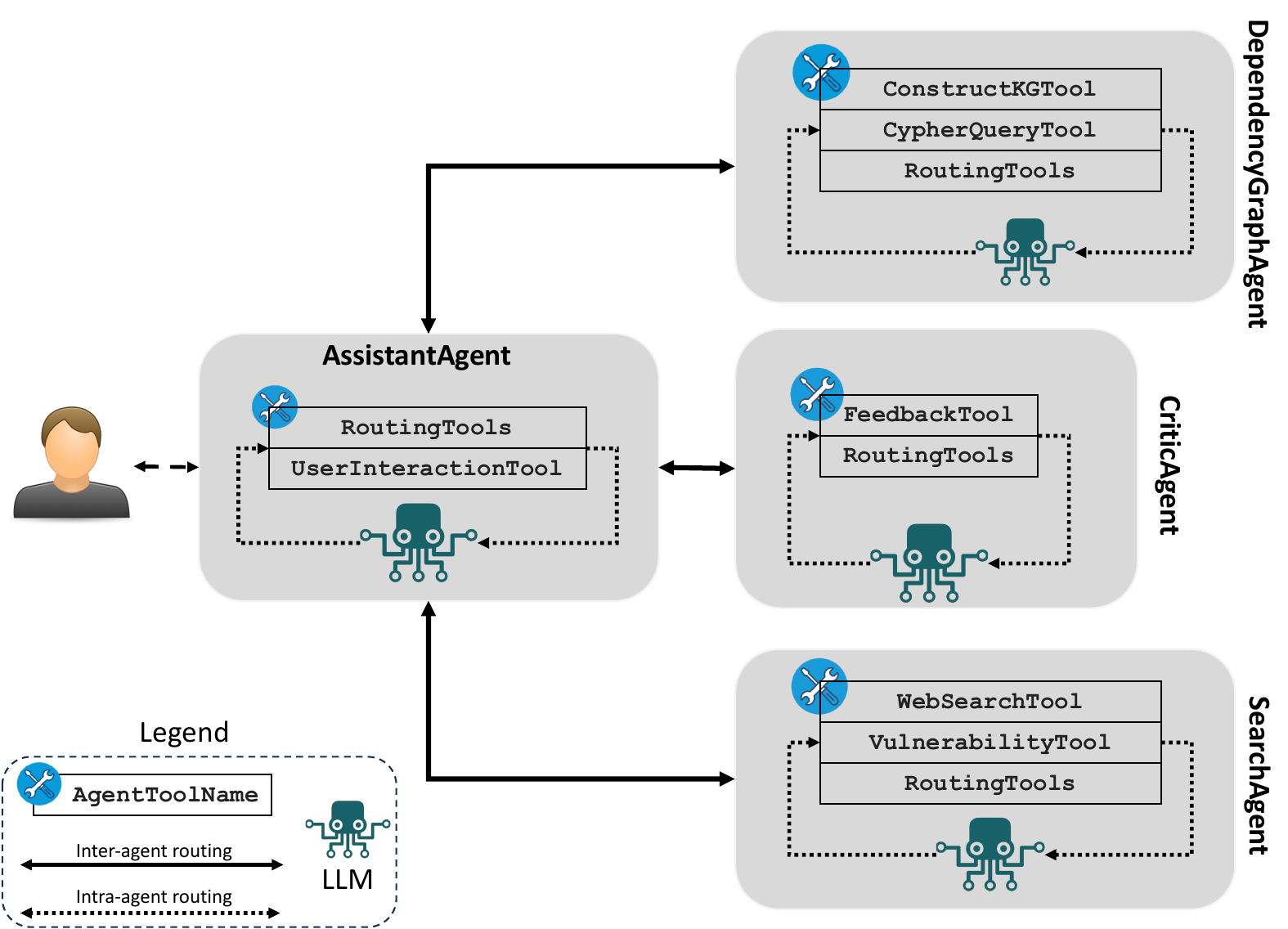}
    \caption{\tool high-level orchestration architecture}
    \label{fig:depRAG_arch}
\end{wrapfigure}

This is the core multi-agent interaction pattern that underlies \tool, and is reminiscent of
Actor/Critic methods in reinforcement learning~\cite{actor-critic}.

In an Agent-Critic framework, the Agent serves as the principal entity responsible for processing external inputs and generating output. It is tasked with achieving a specific goal, is equipped with instructions, and has access to the necessary tools and resources. In our implementation, the agent's objective typically involves specialized question-answering. The available resources may include data sources, other agents, or multi-agent systems, while the tools consist of structured mechanisms to invoke API calls, query databases, or perform computations. The Agent's primary role is to formulate a sequence of queries to these resources in pursuit of its assigned goal. It is required to generate a semi-structured message that includes its proposed solution, the reasoning process leading to this solution, and a justification—citing relevant sources where applicable. This message is then submitted for evaluation and feedback from the Critic. 

To this end, \tool comprises four agents: \assistantagent, \dependencygraphagent, \searchagent, and \criticagent, each designed to enhance the prompt with supplementary information, as illustrated in Figure~\ref{fig:depRAG_arch}. 
Table~\ref{table:toolSummary} lists the tools associated with each agent. 
The \assistantagent coordinates the interactions among these agents by decomposing complex tasks into simpler sub-tasks, delegating them to the appropriate agent, and consulting the \criticagent for feedback on the responses generated before delivering the final answer to the user.
The following is a description of these agents:
\begin{itemize}
    \item \textbf{\dependencygraphagent}. This agent encompasses a suite of tools designed for constructing, visualizing, and retrieving information from the KG. In particular, for retrieving information from the KG, \dependencygraphagent translates user queries expressed in natural language into the query language of the KG, executes the query, and relays the response to \assistantagent. Before accessing the KG, \dependencygraphagent utilizes the \texttt{GraphSchemaTool} to retrieve the KG schema, ensuring the query is generated with the correct entity names and relationships. 
    For tasks such as visualization and schema retrieval, we chose to implement queries directly, bypassing the use of LLMs for query generation. This decision aligns with a key design principle in our system, termed the LLM Minimization Principle. According to this principle, tasks that can be deterministically and explicitly defined within standard programming paradigms should be executed directly, without LLMs, to enhance reliability and minimize token usage and latency~\cite{MALADE}.
    \item \textbf{\searchagent}. This agent is dedicated to searching for information on the Web or in the vulnerability database.
    The Web search tool \texttt{WebSearchTool} aims to provide answers beyond the scope of the KG and vulnerability database. For example, the user could ask \tool about the latest version of a specific package.
    \item \textbf{\assistantagent}. This agent manages the interaction between the user and multiple agents. It further decomposes complex queries into simpler sub-queries, delegates them to the appropriate agents, validates the responses, and aggregates the results into a cohesive answer.
    \item \textbf{\criticagent}. The Critic is an auxiliary agent paired with the primary Agent described above. Its primary function is to evaluate the Agent's reasoning process, ensuring alignment with the given instructions, and provide constructive feedback. This iterative feedback loop has been demonstrated to enhance the quality of LLM-generated outputs~\cite{MALADE,madaan2024self}. The Agent refines its response based on the Critic's input, repeating this process until the Critic approves the output, after which the Agent signals task completion and delivers the final results. 
\end{itemize}

\noindent
% All agents are equipped with a suite of routing tools designed to streamline interactions both between agents and within the tools embedded in each agent, as outlined in Table~\ref{table:toolSummary}. 
In \tool, we developed a suite of advanced routing tools, built on top of the Langroid framework, to enhance reasoning and comprehension within software dependency environments. These extensions are designed to improve the \tool’s ability to handle complex dependency structures and facilitate more accurate analysis and decision-making processes. Specifically, all agents are equipped with a suite of routing tools as outlined in Table~\ref{table:toolSummary}.

% \vspace{-0.3cm}
\section{\tool Proof-of-Concept}
This section describes the implementation of \tool. It then describes the conducted experiments to evaluate \tool.

\subsection{Implementation}
We implemented \tool in Python using Langroid~\cite{langroid}, which is a framework that supports the development of multi-agent LLMs and enables seamless integration with various LLMs. Moreover, Langroid can orchestrate the interaction between agents and contains built-in tools that facilitate the development of RAG applications, such as performing Web search and accessing Neo4j~\cite{neo4j}, a graph database with a declarative query language called Cypher.
To address the orchestration and critic interaction requirements of \tool, we developed supplementary routing tools. For instance, \texttt{QuestionTool} was designed to specifically relay decomposed questions from the \assistantagent to the appropriate agent, as determined by the ``target\_agent'' flag generated within the tool message (see Steps 2 and 3 in Figure~\ref{fig:steps}). While ``ForwardTool'' is a built-in tool in Langroid for routing messages between agents.

\noindent
\textbf{KG Schema.}
The user inputs the package name, version, and ecosystem, after which \tool generates the dependency graph based on the schema depicted in Figure~\ref{fig:schema}. According to this schema, the nodes in the KG belong to the entity Package, which has only two properties, the package name and version. For entity relationships, there is only one relation ``depends\_on''.

\noindent
\textbf{Source of Software Dependencies.}
To obtain the dependencies for the provided project and construct the dependency KG, we used Deps.Dev API,\footnote{\url{https://deps.dev}} a service developed and hosted by Google to help developers understand the structure and security of open-source software packages. This service was used in previous work~\cite{rahman2024characterizing,linuntrustide} to characterize and analyze dependencies. It repeatedly examines websites, such as GitHub and PyPI, to find up-to-date information about open-source software packages, thus generating a comprehensive list of direct and transitive dependencies that are constantly updated. \tool receives a JSON response from the Deps.Dev API that contains a list of dependencies of the provided package. Then ConstructKGTool propagates the data in the JSON data to build the dependency KG.

% we can use similar figures in the paper to illustrate the evaluation https://arxiv.org/pdf/2311.07914. Also, can we use any of the evaluation metrics proposed in this paper?

% https://arxiv.org/pdf/2306.04136, can we do something like table-4

% this paper https://arxiv.org/pdf/2308.03427 evaluates capabilities to generate queries 

% \vspace{-0.2cm}
\subsection{Experiments}
In this study, we address the following research questions (RQs):

\begin{itemize}
    % \item RQ1: Can \tool construct the dependency graph for software packages from various ecosystems?
    \item RQ1: Does \dependencygraphagent generate syntactically and functionally accurate Cypher queries to retrieve information from the dependency KG?
    \item RQ2: Does Agent-Critic interaction, the core design pattern underlying \tool, effectively enhance the reliability of \tool?
\end{itemize}

\noindent
% \textbf{Evaluation Setup.} The experiments were conducted on a system equipped with a 2.3 GHz Quad-Core Intel Core i7 processor, 32 GB of RAM, and running macOS Sonoma. 

We have chosen two representative LLMs to test \tool, summarized in Table~\ref{tab:llms}. 
Both are pre-trained models. GPT-4 Turbo represents a proprietary model, whereas Llama-3 is an open-source model.

\begin{comment}
\subsubsection{RQ1: Software Ecosystems Support}
\tool supports the construction of dependency graphs for four software ecosystems: PyPI, NPM, Cargo, and Go. Figure~\ref{fig:tensorflow} illustrates the dependency graphs of the TensorFlow package in three ecosystems. 

This phenomenon of cross-ecosystem software highlights the dynamic and context-dependent nature of software dependency environments, where a single software package may manifest varying dependency graphs depending on its ecosystem. Capturing this variability is particularly critical in decision-making processes, especially in software translation~\cite{code-translation}, as researchers explore automated methods for translating code across different programming languages. Consequently, \tool can be employed to automate analysis workflows designed to compare the properties of dependency graphs corresponding to the same software package across different ecosystems.
\end{comment}

\subsubsection{RQ1: Accuracy of Cypher Query Generation}
The effectiveness of the agent, \dependencygraphagent, is contingent on the LLM's ability to accurately translate user queries from natural language into Cypher queries for retrieving information from Neo4j database, which stores the dependency knowledge graph (KG). Consequently, it is essential to assess the accuracy of the generated Cypher queries from both syntactical and functional perspectives to ensure reliable performance.

To evaluate this aspect, we asked \tool to answer questions about important properties of graphs (see Table~\ref{tab:questions}) based on the two models selected in this work (listed in Table~\ref{tab:llms}). We conducted this experiment without involving the \criticagent, since according to our implementation, \criticagent provides only a feedback on the final answer generated by \assistantagent.
The questions presented in Table~\ref{tab:questions} are asked after constructing the dependency graph for \texttt{Chainlit} version 1.1.200.

We observed that both LLMs are capable of generating correct Cypher queries, though not consistently. To address this, we implemented a strategy where the LLMs first retrieve the database schema before generating the query and are provided with the error message to prompt a retry if necessary. As shown in Table~\ref{tab:questions} (column 3), Llama-3 failed to produce correct Cypher queries on the first attempt for two questions, requiring one retry for the first question and two for the second. In contrast, GPT-4-Turbo successfully generated accurate Cypher queries on the first attempt for all questions. The final column of Table~\ref{tab:questions} indicates whether the provided answer was correct (without utilizing the \criticagent). Despite Llama-3 producing a syntactically correct Cypher query after two attempts, the resulting query yielded an inaccurate answer. This occurred because the generated query was overly general, returning an overall count of paths without explicitly considering the node \texttt{Chainlit} as the starting point for each path, leading to an irrelevant answer in the context of the dependency graph.

\begin{table}[ht]
\centering
\caption{The performance of \dependencygraphagent to generate accurate Cypher queries based on the selected LLMs after constructing the dependency graph for \texttt{Chainlit} version 1.1.200. \tool is executed without involving the \criticagent.}
\scalebox{0.7}{
\begin{threeparttable}
\begin{tabular}{p{5.7cm}cp{1.9cm}cp{1.5cm}}
\toprule
\textbf{Question} & \textbf{\# of Cypher Query Trials} & \textbf{Model} & \textbf{Correct Response} \\ \midrule
\makecell[l]{What is the depth of the graph?} & 0 & GPT-4-Turbo & Yes \\ \cline{2-4} 
  & 1 & Llama-3 & Yes \\ \midrule
\makecell[l]{Are there cycles in the graph?} & 0 & GPT-4-Turbo & Yes \\ \cline{2-4} 
 & 0 & Llama-3 & Yes \\ \midrule
\makecell[l]{How many path chains are in the graph?\tnote{*}} & 0 & GPT-4-Turbo & Yes \\ \cline{2-4} 
 & 2 & Llama-3 & No \\ 
 %\midrule
% \makecell[l]{What is the density of the graph?} & 2 & 0 & GPT-4-Turbo & Yes \\ \cline{2-5} & 2 & 0 & Llama-3 & No \\ 
 \bottomrule
\end{tabular}
\begin{tablenotes}
\footnotesize
\item[*] Listing~\ref{lst:cypher_queries} presents the queries generated by both Llama-3 and GPT-4-Turbo.
\end{tablenotes}
\end{threeparttable}
}
\label{tab:questions}
\end{table}

\subsubsection{RQ2: Efficiency of the Critic-Agent interaction}

\begin{figure}[htbp]
    \begin{minipage}[b]{0.48\textwidth}
        \centering
        \includegraphics[width=\textwidth]{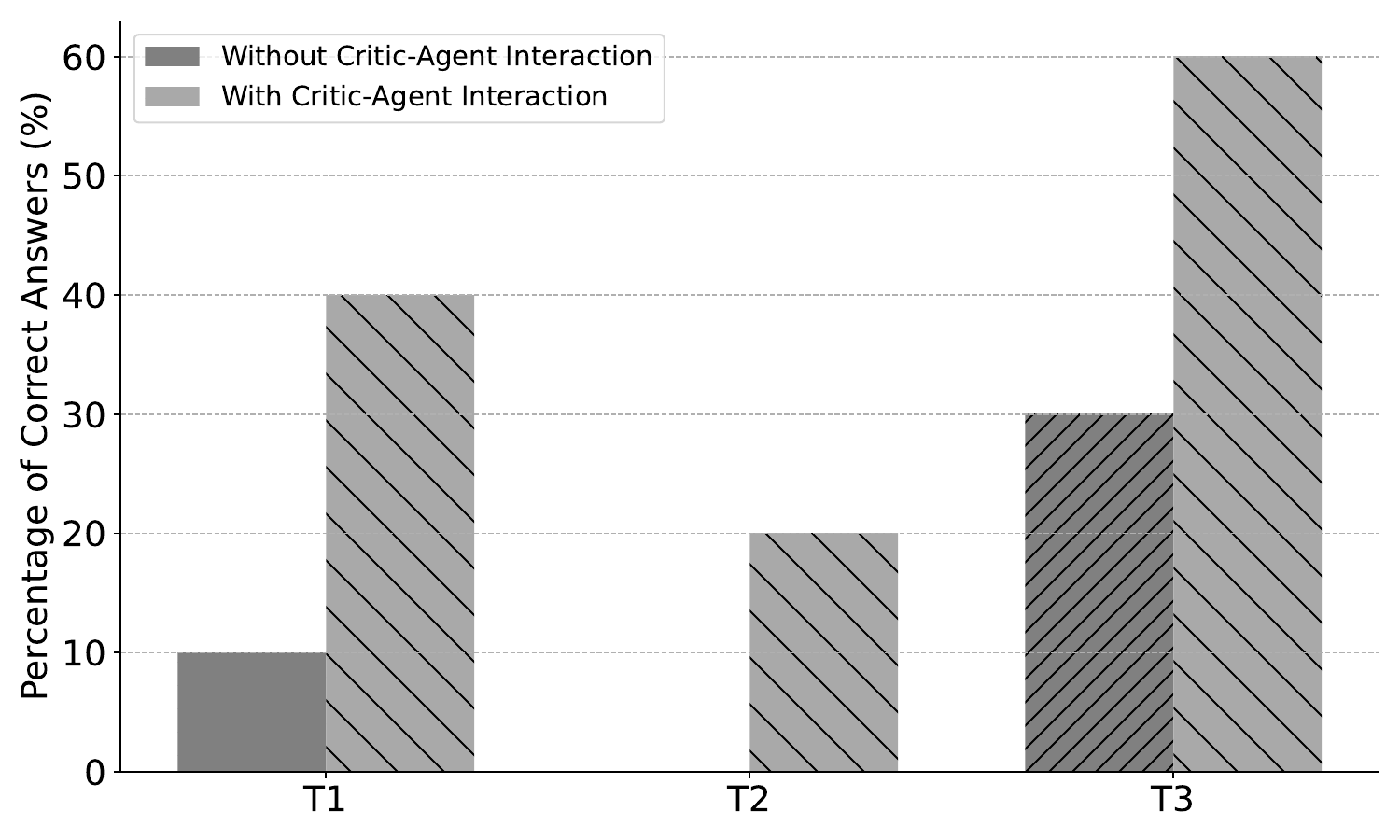}
        \caption{Ablation results show the correctness of the answers with and without Critic-Agent interaction.}
        \label{fig:ablation}
    \end{minipage}
    \hfill
    \begin{minipage}[b]{0.48\textwidth}
        \centering
        \includegraphics[width=\textwidth]{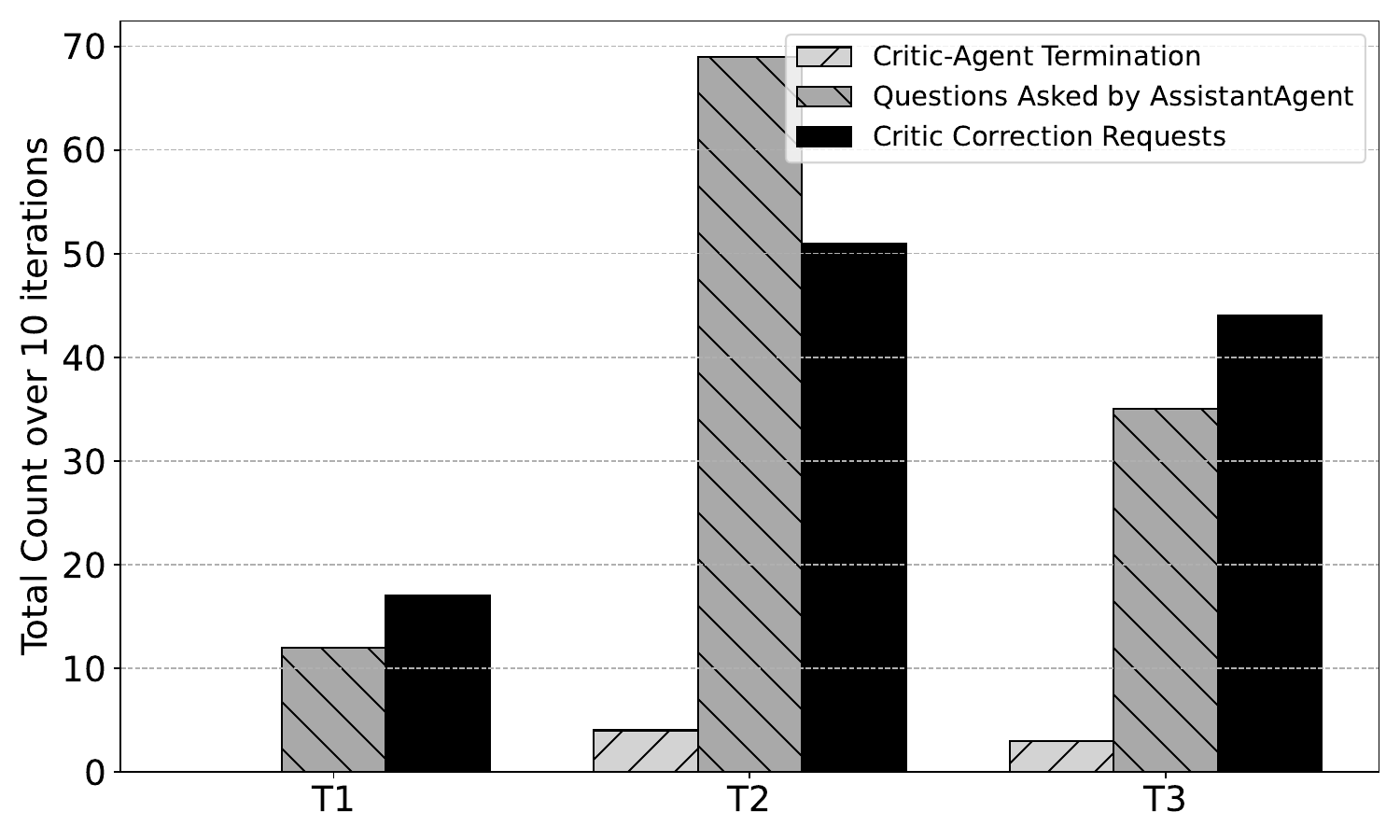}
        \caption{Total number of terminatiom, \assistantagent questions, and \criticagent responses.}
        \label{fig:total_critic}
    \end{minipage}
\end{figure}

We performed an ablation study to assess the effectiveness of the Agent-Critic interaction in \tool. Specifically, we compare the precision of the answers for three tasks (Listing~\ref{lst:tasks}) in ten iterations with and without \criticagent. 
In these experiments, we focus exclusively on GPT-4-Turbo, as our preliminary observations indicated that Llama-3 encountered difficulties following the orchestration and Critic interaction outlined in Figure~\ref{fig:depRAG_arch}. We restricted the Critic-Agent interaction to the final answer generated by the \assistantagent. This decision was made to minimize the number of iterative exchanges with other agents, as increasing these iterations would negatively impact performance by increasing token costs and runtime, while also reducing reliability~\cite{MALADE,repair-alloy}. 
The results are presented in Figure~\ref{fig:ablation}. We manually verified the accuracy of the answers in both experiments and observed that the Critic-Agent interaction improves the quality and correctness of the answers. Specifically, the ablation results significantly improved accuracy with the Critic-Agent mechanism. Without \criticagent, the precision was $13.3\%$, rising to $40\%$ after its integration, highlighting the impact of \criticagent on the accuracy of the response. Subsequently, introducing the critic agent resulted in the \tool being \textbf{three times} more accurate.

We extend our analysis of the Agent-Critic interaction by examining the frequency of \criticagent interventions to correct errors in the responses generated by \assistantagent. Instances where \criticagent provides feedback, are characterized by multiple rounds of interaction between the \assistantagent and \criticagent. The findings of this analysis, including the total number of terminations, are illustrated in the histogram in Figure~\ref{fig:total_critic}. In certain cases, we observe unproductive exchange cycles between \criticagent and \assistantagent, where no progress is achieved. To address this, we impose a limit of ten feedback iterations per interaction, after which the session is terminated, and the final response is recorded. We also tracked the total number of questions posed by the \assistantagent, as shown in Figure~\ref{fig:total_critic}. The data reveal that $T2$ involved a higher number of questions, which is attributed to the \assistantagent querying the vulnerability of each identified package individually. This increase in question count is quantified by monitoring the use of the \texttt{QuestionTool}.

% This is vital for breaking tasks into manageable sub-queries because LLMs often struggle with multi-step reasoning. To evaluate this aspect, we initially asked both models: ``tell me two interesting things about the dependency graph.'' Both demonstrated the capability to generate multiple Cypher queries, receive the answers and provide a complete answer. Table~\ref{tab:questions} provides examples for user questions and the corresponding Cypher queries, along with the responses generated by the considered model. The last question provides an additional example to demonstrate the capability of \tool to perform task planning. The computation of graph density involves breaking this task into two steps: (1) retrieving the number of nodes (N) and edges (E) and (2) calculating the density according to this formula $\frac{E}{N(N-1)}$, since the KG is a directed graph. Llama-3 performed both steps but failed to give a correct answer as it assumed that the KG is an undirected graph and applied the incorrect formula.

% \vspace{-0.35cm}
\section{Related Work}

% \textbf{Multi-agent RAG.}

\noindent
\textbf{Critic and Feedback in LLM Apps.}
Konda et al.~\cite{actor-critic} presented the Actor-Critic Algorithm, which is a type of Reinforcement Learning (RL) algorithm that combines aspects of both policy-based methods (i.e., actor) and value-based methods (i.e., critic). This hybrid approach is designed to address the limitations of each method when used individually. In our approach, we adapt the CRITIC framework proposed by Gou et al.~\cite{gou2023critic} to reinforce LLM agents not by updating weights, which is the case in RL, but through linguistic feedback from critic agents in a reflective, feedback loop.
Alhanahnah et al.~\cite{repair-alloy} present a multi-agent tool for repairing Alloy specifications, with feedback agents but lacking retrieval agents like \dependencygraphagent and \searchagent, which are used in \tool.
MALADE~\cite{MALADE}, a multi-agent system for answering medication side-effect queries, employs Critic-Agent interaction and retrieves data from unstructured documents. \tool, by contrast, primarily uses a knowledge graph and the Web for retrieval.

\noindent
\textbf{Software representation as KG.}
Litzenberger et al.~\cite{DGMF} proposed a unified data model to implement and construct dependency graphs for arbitrary repositories, thus facilitating the comparison of dependency graphs between different repositories. In their implementation, the data model is based on Neo4j, which makes is compliant with \tool's KG schema. Maninger et al.~\cite{code_kg} proposed a visionary approach involving KGs to create a trustworthy AI software development assistant. Specifically, KGs can enable LLMs to correctly and appropriately explain the generated code. To our knowledge, we are the first to investigate the ability of LLMs to generate KGs for software dependencies, and utilize these graphs to aid in responding to queries concerning the dependency structure.
Musco et al.~\cite{musco2017generative} constructed dependency graphs for Java programs at the class level to understand commonalities between different Java projects. In contrast, \tool supports four software ecosystems and constructs dependency graphs at the package level.

% Bommarito et al.~\cite{bommarito2019empirical} analyzed PyPI, including both package metadata and package source, covering software dependencies, authors, licenses, and other summary information over time. 

% \cite{api_recommendation}
% PyEGo~\cite{PyEGo}
% \cite{DGDB}
% \cite{graph4code}
% \vspace{-0.25cm}
\section{Conclusion and Future Work}

We presented \tool, the first steps of an agent-based approach for automated planning and reasoning about software dependencies. Our evaluation showed that Critic-Agent interaction enhances the quality, correctness, and reasoning of LLM responses. However, implementing this architecture necessitates efficient orchestration and mechanisms for preventing feedback loops.
Future work will explore several key directions:
    
\begin{itemize}
    \item \textbf{Evaluating different Critic-Agent Settings.} In this work, we designed and implemented \criticagent to provide feedback on the final answer generated by \assistantagent, which integrates responses from the retriever agents (\dependencygraphagent and \searchagent). However, it is crucial to assess various scenarios to optimize the effectiveness of Critic-Agent interactions.
    % \item \textbf{Extending Dependencies.} The current KG is constructed based on Deps.Dev API that suffers from some limitations. For example, Python allows defining optional dependencies, indicating that these dependencies will not be installed by default. However, the Deps.Dev API does not capture these optional dependencies. This lack of support can lead to serious challenges and conflicts after updating packages that depend on optional packages. Therefore, retrieving dependencies from code and other dependency configuration files, such as ``pyproject.toml,'' is crucial. Moreover, certain software ecosystems may incorporate dependencies originating from different ecosystems. Previous studies have demonstrated the widespread utilization of C libraries within Python and NPM projects~\cite{insight}. Consequently, the dependency KG can be expanded to encompass these relationships, distinguishing between inter-ecosystem and intra-ecosystem dependencies.
    % 44\% view SBOMs as essential for managing vulnerabilities, one primary concern in today’s software development environments [22]. (from SBOM review paper)
    \item \textbf{Generating Software Bill of Materials (SBOM).} The generated dependency graph encompasses both direct and transitive dependencies, enhancing the suitability of \tool for SBOM generation and providing metadata such as path chains (that is, dependency hierarchy~\cite{sbom}), as opposed to presenting a flat list of dependencies. SBOM format specifications, such as CycloneDX~\cite{cycloneDX}, are designed to accommodate this form of structured data. The hierarchical structure of the dependencies upheld by \tool is in accordance with the minimum SBOM requirements prescribed by the US National Telecommunications and Information Administration (NTIA). \tool exceeds these regulatory requirements, which require disclosure of primary dependencies (first-level) and all ensuing transitive dependencies (second-level)~\cite{ntia}, by documenting the complete chain of dependencies. The generation of SBOM will enable \tool to support vulnerability management, which requires designing a dedicated retrieval to access different sources of vulnerability databases.
    \item \textbf{Dependency Resolution.} Updating dependencies is cumbersome and can introduce incompatibilities that break the application~\cite{upcy}. \tool maintains knowledge about all dependencies within the application, thus qualifying it to provide recommendations that suggest non-vulnerable package versions and will not lead to compatibility issues. 
\end{itemize}

\begin{ack}
We would like to thank Benoit Baudry for the helpful discussions and feedback.
This material is based on work supported by the Office of Naval Research (ONR) under Contract N00014-24-1-2049. Any opinions, findings, conclusions, or recommendations expressed in this material are those of the authors and do not necessarily reflect the views of ONR.
\end{ack}

\medskip
{
\small
\bibliographystyle{plain}
\bibliography{ref,kg}
}
\clearpage
\appendix
\section{Appendix / supplemental material}

\begin{table}[h]
\centering
\caption{Summary of tools used by \tool Agents.}
\scalebox{0.68}{
\begin{tabular}{lll}
\toprule
\textbf{Tool Name} & \textbf{Corresponding Agent} & \textbf{Purpose} \\ \midrule
ConstructKGTool & \dependencygraphagent & Construct the KG \\
GraphSchemaTool & \dependencygraphagent & Obtain the KG schema \\ 
CypherQueryTool & \dependencygraphagent & \begin{tabular}[c]{@{}l@{}}Receive the generated query,\\execute it,  and return the result\end{tabular} \\ 
VisualizeKGTool & \dependencygraphagent & Visualize the KG \\
WebSearchTool & \searchagent & \begin{tabular}[c]{@{}l@{}}Perform Web search and return the\\ most relevant\end{tabular} \\ 
VulnerabilityTool & \searchagent & \begin{tabular}[c]{@{}l@{}}Search for vulnerabilities on OSV\\vulnerability database \end{tabular} \\ 
UserInteractionTool & \assistantagent & \begin{tabular}[c]{@{}l@{}}Control the user interaction with\\the agents\end{tabular} \\
QuestionTool & All except \criticagent & Inter-agent routing \\
ForwardTool & \assistantagent and \criticagent & Inter-agent routing \\
\bottomrule
\end{tabular}
}
\label{table:toolSummary}
\end{table}

\begin{table}[ht]
\centering
\caption{Characteristics of the selected LLMs}
% \vspace{-0.4cm}
\scalebox{0.67}{
\begin{tabular}{llllll}
\toprule
\textbf{Model} & \textbf{Version} & \textbf{Cut-off} & \textbf{\begin{tabular}[c]{@{}l@{}}Context Window\\ (Tokens)\end{tabular}} & \begin{tabular}[c]{@{}l@{}}Input Cost\\ per 1M tokens\end{tabular} & \begin{tabular}[c]{@{}l@{}}Output Cost\\ per 1M tokens\end{tabular} \\ \midrule
GPT-4 Turbo & 1106-preview & Apr 2023 & 128k & \$10 & \$30 \\ 
Llama-3 & 70b-instruct & Dec 2023 & 8k & N/A & N/A \\ \bottomrule
\end{tabular}
}
\label{tab:llms}
\end{table}

\begin{figure*}[h]
    \centering    \includegraphics[width=\textwidth]{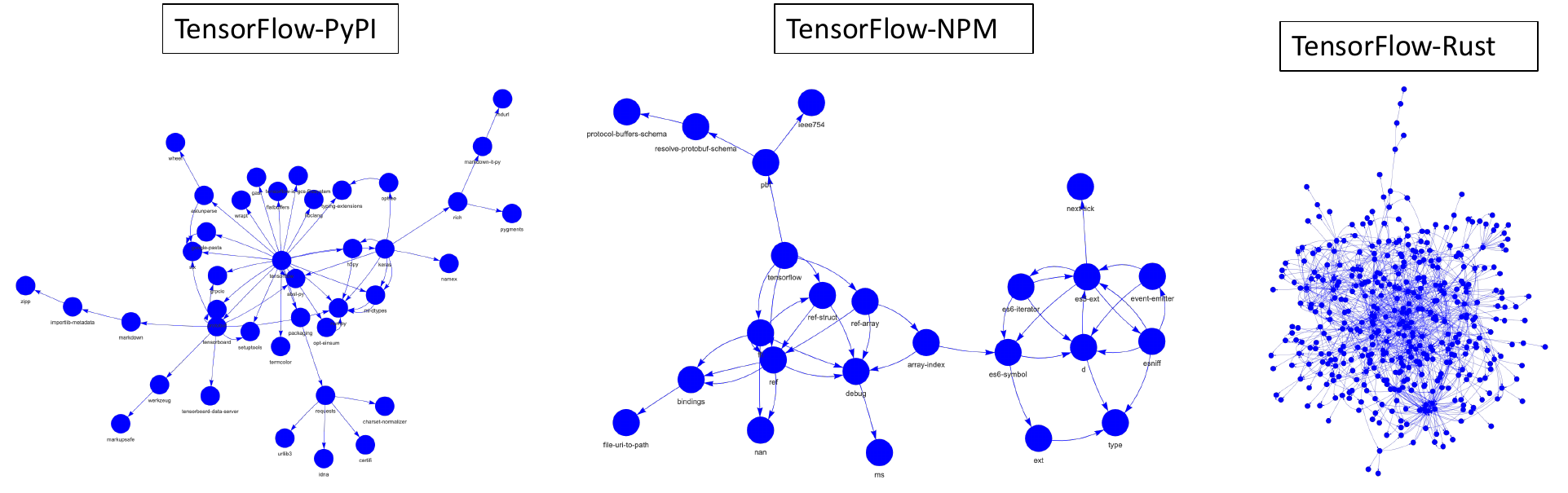}
    \vspace{-0.2cm}
    \caption[Dependency graphs generated by \tool for TensorFlow]{Dependency graphs generated by \tool for TensorFlow across 3 ecosystems (PyPI, NPM, and Rust). Versions 2.16.1, 0.7.0, and 0.21.0, respectively.}
    \label{fig:tensorflow}
\end{figure*}

\begin{figure}[h!]
    \centering   \includegraphics[width=0.44\linewidth,keepaspectratio]{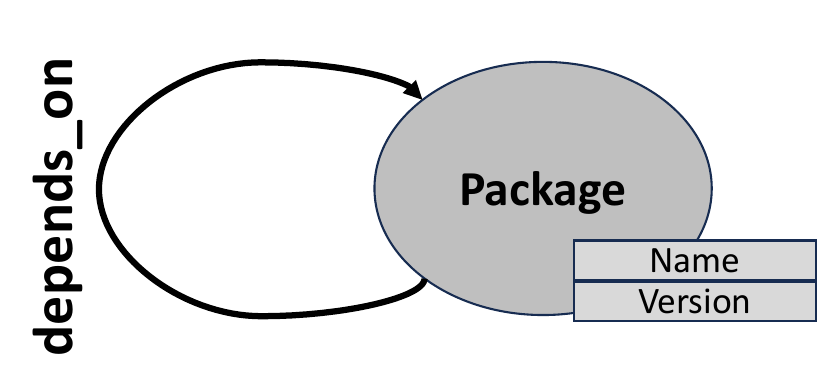}
    \caption{Dependency KG Schema consisting of a single entity with two attributes, ``name'' and ``version'', and one relation.}
    \label{fig:schema}
\end{figure}

\begin{lstlisting}[language=SQL, caption={Cypher queries generated by both models translating question 3 in Table~\ref{tab:questions}},label={lst:cypher_queries}]
-- Query generated by Llama-3
MATCH p=()-[r:DEPENDS_ON*]->() RETURN count(p) 
AS pathCount

-- Query generated by GPT-4-Turbo
MATCH p=(root:Package {name: 'chainlit', version: '1.1.200'})-[:DEPENDS_ON*]->(leaf:Package) 
WHERE NOT (leaf)-[:DEPENDS_ON]->() RETURN count(p) 
AS pathCount
\end{lstlisting}
%Llama-3 result: [{"pathCount": 655}]
%GPT-4-Turbo result: [{"pathCount": 133}]

\lstset{
    basicstyle=\ttfamily,
    breaklines=true,
    frame=single,
    language=Python,  % or just for formatting purposes
    showstringspaces=false
}

\begin{lstlisting}[caption={List of tasks used in RQ3}, label={lst:tasks}]
{
    "T1": "what's the density of the dependency graph of chainlit version 1.1.200 pypi",
    
    "T2": "which packages in chainlit version 1.1.200 pypi have the most dependencies relying on them (i.e., nodes have the highest in-degree in the graph), and what is the risk associated with a vulnerability in those packages?",
    
    "T3": "In the dependency graph of chainlit version 1.1.200 pypi, are there any multi-version conflicts where different packages depend on different versions of the same package? If yes, provide examples of these conflicts and all paths that lead to these packages from the root node"
}
\end{lstlisting}
\end{document}